# About the Justification of Experience Rating: Bonus Malus System and a new Poisson Mixture Model

Magda Schiegl

Cologne University of Applied Sciences
Claudiusstr. 1, D-50678 Cologne
magda.schiegl@fh-koeln.de

**Abstract**

The claim experience of the past is a very important information to calculate the fair price of an insurance contract. In a lot of European countries for instance the prices for motor car insurance depend on the number of claims the driver has reported to the insurance company during the last years. Classically these prices are calculated on the basis of a mixed Poisson model with a gamma mixing distribution. The mixing distribution models the car drivers' qualities across the insured portfolio. This is just one example for experience rating. In the classical context the price is equal to the expectation of the Bayesian posterior distribution.

In some lines of business (especially third party liability and lines with exposure to extreme weather events) we that the real world data cannot be described well enough by the classical Poisson – gamma model. Therefore we investigate the influence of the mixing distribution on the posterior distribution conditional on the experienced number of claims. This enables the application of other – more risk adequate premium principles than the expectation principle. We introduce the inverse – gamma distribution as a new mixing distribution to model claim numbers and compare it to the classical gamma distribution. In both cases a closed analytic representation of the mixed distribution can be found: In the classic case the well known negative binomial distribution, in our new one a representation using the Bessel functions. Additionally we present numerical results about the tail behaviour of the mixed Poisson – inverse – gamma distribution.
Finally we introduce the concept of resolution. It enables us to decide if the classification of risk groups via the number of experienced claims is a risk adequate procedure.

# 1. Introduction

The application of the classical Poisson-gamma theory (see for instance Mack 1997, chap. 2.5.2.) to non – automobile lines of business delivers results that seem wrong to practitioners in some cases: The penalising of experienced claims is rather strict in some examples and too weak in others. This effect occurs especially in those lines of business where an appropriate volume measure for the insured risk is not the number of contracts (as in automobile) but others as for example the insured sum. Insurance portfolios often consist of contracts with a rather wide range of different volumes. This adds an additional source of inhomogeneity to the claim frequency distribution. In those cases we find, that the empirical claim number distribution arising from the claim portfolio does not fit well to the negative binomial (posterior) distribution of the classical model. In practice we often find claim number distributions with heavy tails. Therefore we look at the Bayes model level for a mixing distribution that gives rise to a heavy tailed (posterior) mixed distribution.
The paper of Frangos and Vrontos (2001) shows that an exponential – mixture of the inverse – gamma distribution generates a Pareto distribution. Transferring this idea to the discrete case we investigated the mixed Poisson – inverse – gamma distribution. We report our results in this paper and compare them to those of the classic Poisson – gamma distribution.

Pricing in the framework of classic Bonus – Malus System (BMS) applies the expectation value principle and sets the conditional (i.e. the claim experience) expectation equal to the a posteriori premium. We are interested in the conditional distributions behind and not only in their first moment: What are the effects on pricing? This question leads us to the concept of resolution introduced in section 2.3. and to the insight that not only the expectation but also the width of the conditional distributions is important for experience rating / pricing in insurance practice. The application of a premium principle extracts one number out of a distribution for each risk group – the fair price. For premium principles in general see for instance Schmidt (2002) or Heilmann (1989).
In the literature numerous premium principles and many different mixed Poisson distributions are investigated. Some examples together with the used claim frequency distribution are: Lemaire (1995) (negative binomial), Tremblay (1992) (Poisson inverse Gauß), Walhin and Paris (1999) (Hofmann). Overviews on Bonus Malus systems can be found in the introductions of the following papers: Frangos and Vrontos (2001), Heras et al. (2004), Pitrebois et al. (2005). General Papers on BMS are: Bühlmann (1967), Bühlmann and Gisler (1997), Jewell and Schnieper (1985), Pinquet (1998), Schnieper (1995), Taylor (1997).
To our knowledge the results of the inverse gamma distribution as mixing distribution have never been reported in the literature. We find in this case analytical, closed form solutions for the relevant distributions and for their first two moments.
The paper is organised as follows: In section 2 a very short summary of Bayes theory is given and the two investigated mixed Poisson models, the Poisson gamma (2.1.) and the Poisson inverse gamma (2.2) are introduced. An excursion on the tail dependence of the mixed Poisson inverse gamma distribution is given in 2.2.1. In section 2.3. the concept of resolution is introduced. In section 3 we finally summarize our results and give an outlook.

## 2. Experience rating – Definitions and Bayes formulas

In the following we introduce the relevant quantities that are analysed in this paper. The mixed Poisson distribution is defined as:

$$P(N = n) = \int_0^\infty \vartheta^n e^{-\vartheta} f(\vartheta) d\vartheta / n! \qquad (2.0.1)$$

$$\text{for } n = 0,1,\ldots \text{ and } \vartheta > 0$$

The idea behind this formula is that the claim numbers are generated by an inhomogeneous insurance portfolio. Therefore the Poisson parameter $\vartheta$ is not constant within the portfolio but varies from risk to risk and is distributed according to the function $f(\vartheta)$. The higher the value of the Poisson parameter the higher is the risk of the individual insured object. We analyse in this paper two different types of mixing distributions: The gamma distribution and the inverse – gamma distribution. The above formula is valid for an observation period of one year and n is the observed number of claims (of an individual risk) in one year. This time scale (one year) can be changed easily in this case as the convolution of several Poisson distributions is again a Poisson distribution whose parameter is the sum of the convolved distributions' parameters. We generalise for J years – so the mixed distribution reads:

$$P(N = n) = \int_0^\infty (J\vartheta)^n e^{-J\vartheta} f(\vartheta) d\vartheta / n! \qquad (2.0.2)$$

We assume implicitly that there is no change during the J years in the risk situation or - technically speaking - in the distribution parameters.
Expectation and variance of the independent distribution (2.0.2) are related with the expectation and variance of the dependent distribution in the following way:

$$E(N) = E(E(N|\Theta)) = J \cdot E(\Theta)$$
$$Var(N) = E(Var(N|\Theta)) + Var(E(N|\Theta)) = J \cdot E(\Theta) + J^2 \cdot Var(\Theta)) \qquad (2.0.3)$$

Via the Bayes formula the dependent and independent distributions are related:

$$P(\Theta = \vartheta | N = n) = \frac{P(N = n | \Theta = \vartheta) \cdot P(\Theta = \vartheta)}{P(N = n)} \qquad (2.0.4)$$

In the classical BMS the a posteriori premium is set equal to the dependent expectation: $E(\Theta | N = n)$ (Bichsel 1964; Mack 1997, chap. 2.5.2.).
So far the classical BM theory – we now are interested in the dependent distribution of claim numbers in a time interval t1 < t < t2 under the condition that the number of claims in an earlier time interval t0 < t < t1 is known. Let the lengths of the two intervals be $J_1 = t_1 - t_0$ and $J_2 = t_2 - t_1$ the random variable of number of claims in the first interval is $N_1$, $N_2$ in the second . We are interested in the distribution density

$$P(N_2 = n_2 | N_1 = n_1) = \int P(N_2 = n_2 | \Theta = \vartheta) \cdot P(\Theta = \vartheta | N_1 = n_1) d\Theta =$$
$$\int \frac{1}{n_2!}(J_2\Theta)^{n_2} e^{-J_2\Theta} P(\Theta = \vartheta | N_1 = n_1) d\vartheta \qquad (2.0.5)$$

for $n_1 = 0,1,...$, $n_2 = 0,1,...$ and $J_1, J_2 > 0$

This is again a mixed Poisson distribution with mixing distribution $P(\Theta = \vartheta | N_1 = n_1)$ (2.0.4). According to (2.0.3 and 2.0.2) the conditional mean (i.e. the a posteriori expectation) and the variance of $N_2$ conditional on $N_1$ are given by:

$$E(N_2 | N_1) = J_2 \cdot E(\Theta | N_1)$$
$$Var(N_2 | N_1) = J_2 \cdot E(\Theta | N_1) + J_2^2 \cdot Var(\Theta | N_1) \qquad (2.0.6)$$

## 2.1. The Poisson – gamma model

In the framework of the classical Poisson-gamma model the above defined quantities with the mixing gamma distribution (see Mack 1997, chap. 2.5.2.)

$$f(\vartheta) = \beta^\alpha \vartheta^{\alpha-1} e^{-\beta\vartheta} / \Gamma(\alpha) \qquad (2.1.1)$$

The above defined quantities evaluate as (see 2.0.2):

$$P(N = n) = \binom{n+\alpha-1}{n} p^\alpha (1-p)^n \qquad (2.1.2)$$

with

$$p = \frac{\beta}{\beta + J} \qquad \text{for } n = 0,1,... \text{ and } \alpha, p > 0$$

This is the negative binomial distribution – a well known result.
The unconditional expectation and variance according to (2.0.3) are:

$$E(N) = J\alpha/\beta$$
$$Var(N) = J\alpha/\beta + J^2\alpha/\beta^2 = (1 + \frac{J}{\beta})E(N) \qquad (2.1.3)$$

The conditional pdf is a gamma distribution in this case and according to (2.0.4) it reads:

$$P(\Theta = \vartheta \mid N = n) = \frac{e^{-J\vartheta}(J\vartheta)^n}{n!} \frac{\beta^\alpha \vartheta^{\alpha-1} e^{-\beta\vartheta}}{\Gamma(\alpha)} \Bigg/ \frac{J^n \beta^\alpha \Gamma(n+\alpha)}{n! \Gamma(\alpha)(\beta+J)^{n+\alpha}}$$
$$= (\beta+J)^{\alpha+n} \vartheta^{\alpha+n-1} e^{-(\beta+J)\vartheta} \Big/ \Gamma(\alpha+n) \tag{2.1.4}$$

The dependent expectation and variance are received from distribution (2.1.4) and evaluate as:

$$E(\Theta \mid N = n) = \frac{\alpha+n}{\beta+J} = \frac{J}{\beta+J}\frac{n}{J} + \frac{\beta}{\beta+J}\frac{\alpha}{\beta}$$
$$Var(\Theta \mid N = n) = \frac{\alpha+n}{(\beta+J)^2} \tag{2.1.5}$$

As is well known (see Bühlmann 2005, Chap. 2.4) within the Poisson – gamma – model the dependent expectation can be written in a credibility form as is done above -- with the credibility factor $\frac{J}{\beta+J}$ weighting the experienced claims on the one hand and the portfolio mean claim number on the other hand. Notice: As J is the number of observation years the two distribution parameters can be interpreted as follows: β is the number of years an individual risk has to be observed in order to have equal credibility weights for the experienced claims on the one hand and the portfolio mean on the other. The Parameter α can be interpreted as the expected number of claims occurring in a time period with length β in the portfolio mean.

According to (2.0.5) and (2.1.4) the conditional claim number distribution evaluates as a negative binomial distribution:

$$P(N_2 = n_2 \mid N_1 = n_1) = \binom{n_2 + n_1 + \alpha - 1}{n_2} p^{\alpha+n_1}(1-p)^{n_2} \tag{2.1.6}$$

$$\text{with} \quad p = \frac{\beta + J_1}{\beta + J_1 + J_2}$$

$$\text{for } n_1 = 0,1,\ldots,\ n_2 = 0,1,\ldots \text{ and } J_1, J_2 > 0$$

It is a mixed Poisson distribution (see 2.0.5) with the gamma distribution (2.1.4) as mixing distribution and with the conditional and unconditional expectations and variances (see (2.0.6.)):

$$E(N_2|N_1=n_1) = J_2 \frac{n_1+\alpha}{\beta+J_1}$$

$$Var(N_2|N_1=n_1) = J_2 \left(\frac{n_1+\alpha}{\beta+J_1}\right)\left(\frac{\beta+J_1+J_2}{\beta+J_1}\right) \quad (2.1.7)$$

(see (2.1.3))

$$E(N_2) = J_2 \frac{\alpha}{\beta}$$

$$Var(N_2) = J_2 \frac{\alpha}{\beta} + J_2^2 \frac{\alpha}{\beta^2} \quad (2.1.8)$$

## 2. 2. The Poisson – inverse – gamma model

We additionally investigate another Poisson mixture model: The Poisson - inverse - gamma model with the mixing pdf being the inverse – gamma distribution:

$$f(\vartheta) = \frac{\frac{1}{m} \cdot e^{-\frac{m}{\vartheta}}}{\left(\frac{\vartheta}{m}\right)^{s+1} \cdot \Gamma(s)} \quad (2.2.1)$$

The mixed pdf has two parameters: m and s. Its mean is $\frac{m}{s-1}$ and its variance $\left(\frac{m}{s-1}\right)^2 \frac{m}{s-2}$.

The resulting unconditional distribution according to (2.0.2), (2.2.1) and (2.2.4) reads:

$$\begin{aligned} P(N=n) &= \frac{J^n m^s}{n!\Gamma(s)} \int_0^\infty \exp\left(-J\vartheta - \frac{m}{\vartheta}\right) \vartheta^{n-s-1} d\vartheta \\ &= 2\frac{J^n m^s}{n!\Gamma(s)} \left(\frac{J}{m}\right)^{-\frac{n-s}{2}} K_{n-s}\left(2\sqrt{Jm}\right) \\ &= \frac{2(Jm)^{\frac{s+n}{2}}}{n!\Gamma(s)} K_{s-n}(2\sqrt{Jm}) \end{aligned} \quad (2.2.2)$$

$$\text{for } n = 0,1,\ldots \text{ and } m, s > 0$$

with the Bessel function of the third kind defined as:

$$K_n(x) = \frac{1}{2}\int_0^\infty \exp\left(-\frac{1}{2}x\left(y+\frac{1}{y}\right)\right) y^{n-1} dy, \qquad x > 0 \qquad (2.2.3)$$

Due to this definition the following integral can be written as:

$$\int_0^\infty \exp\left(-jx - \frac{a}{x}\right) \cdot x^b dx = 2\left(\frac{j}{a}\right)^{-(1+b)/2} K_{1+b}\left(2\sqrt{aj}\right) \qquad (2.2.4)$$

This relation has been used in the derivation of the unconditional distribution (2.2.2) and will be used further on.
According to (2.0.3) and (2.2.1) the unconditional expectation and variance are:

$$E(N) = \frac{Jm}{s-1}$$

$$Var(N) = \frac{Jm}{s-1} + \left(\frac{Jm}{s-1}\right)^2 \frac{1}{s-2} \qquad (2.2.5)$$

The conditional pdf can be calculated by inserting (2.2.1) and (2.2.2) into (2.0.4):

$$P(\Theta = \vartheta \mid N = n) = \frac{\dfrac{(J\vartheta)^n}{n!}e^{-J\vartheta} \cdot \dfrac{\dfrac{1}{m}e^{-\frac{m}{\vartheta}}}{\left(\dfrac{\vartheta}{m}\right)^{s+1}\Gamma(s)}}{\dfrac{2(Jm)^{(s+n)/2}}{n!\Gamma(s)}K_{s-n}\left(2\sqrt{Jm}\right)} = \frac{e^{-\left(J\vartheta+\frac{m}{\vartheta}\right)}\vartheta^{n-s-1}\left(\dfrac{m}{J}\right)^{\frac{s-n}{2}}}{2K_{s-n}\left(2\sqrt{Jm}\right)} \qquad (2.2.6)$$

The conditional expectation and variance can be evaluated under consideration of (2.2.4):

$$E(\Theta \mid N = n) = \int_0^\infty \vartheta \cdot P(\vartheta\mid n)d\vartheta = \sqrt{\frac{m}{J}}\frac{K_{s-n-1}\left(2\sqrt{Jm}\right)}{K_{s-n}\left(2\sqrt{Jm}\right)}$$

$$Var(\Theta \mid N = n) = \int_0^\infty \vartheta^2 \cdot P(\vartheta\mid n)d\vartheta - E(\vartheta\mid n)^2 = \qquad (2.2.7)$$

$$= \frac{m}{J}\left[\frac{K_{s-n-2}\left(2\sqrt{Jm}\right)}{K_{s-n}\left(2\sqrt{Jm}\right)} - \left(\frac{K_{s-n-1}\left(2\sqrt{Jm}\right)}{K_{s-n}\left(2\sqrt{Jm}\right)}\right)^2\right]$$

According to (2.0.5) and (2.2.4) the conditional claim number pdf evaluates as:

$$P(N_2 = n_2 | N_1 = n_1) =$$

$$\frac{J_2^{n_2}}{n_2!} \left(\frac{m}{J_1}\right)^{\frac{s-n_1}{2}} \left(\frac{J_1 + J_2}{m}\right)^{\frac{s-n_1-n_2}{2}} \frac{K_{s-n_1-n_2}\left(2\sqrt{m(J_1+J_2)}\right)}{K_{s-n_1}\left(2\sqrt{mJ_1}\right)} \qquad (2.2.8)$$

$$\text{for } n_1 = 0,1,\ldots,\ n_2 = 0,1,\ldots \text{ and } J_1, J_2 > 0$$

with the conditional and unconditional expectations and variances of the mixed Poisson distribution ( see 2.0.3 and 2.2.7):

$$E(N_2 | N_1 = n_1) = J_2 \sqrt{\frac{m}{J_1}} \frac{K_{s-n_1-1}\left(2\sqrt{J_1 m}\right)}{K_{s-n_1}\left(2\sqrt{J_1 m}\right)}$$

$$Var(N_2 | N_1 = n_1) =$$

$$J_2 \sqrt{\frac{m}{J_1}} \frac{K_{s-n_1-1}\left(2\sqrt{J_1 m}\right)}{K_{s-n_1}\left(2\sqrt{J_1 m}\right)} + J_2^2 \frac{m}{J_1} \left[\frac{K_{s-n_1-2}\left(2\sqrt{J_1 m}\right)}{K_{s-n_1}\left(2\sqrt{J_1 m}\right)} - \left(\frac{K_{s-n_1-1}\left(2\sqrt{J_1 m}\right)}{K_{s-n_1}\left(2\sqrt{J_1 m}\right)}\right)^2\right] \qquad (2.2.9)$$

see (2.2.5)

$$E(N_2) = J_2 \frac{m}{s-1}$$

$$Var(N_2) = J_2 \frac{m}{s-1} + \left(\frac{J_2 m}{s-1}\right)^2 \frac{1}{s-2} \qquad (2.2.10)$$

### 2. 2. 1. The Tail Behaviour

We investigate the <u>tail behaviour</u> of the mixed Poisson – inverse gamma distribution (see eq. 2.2.2.). In order to learn about this subject we transform the number density in the following way:

$$y := \ln(P(N = n));$$
$$x := \ln(n)$$

If the distribution has a fat tail then y and x show linear dependence for large x. Two examples: For the Poisson distribution we find a clear non linear dependence:

$$y = \bar{c}_0 + \bar{c}_1 \cdot e^x - \left(e^x + \frac{1}{2}\right) \cdot x$$

and analogously for the negative binomial (= mixed Poisson – gamma) distribution:

$$y = c_0 + c_1 \cdot e^x + (c_2 + e^x) \cdot \ln(c_3 + e^x) - \left(e^x + \frac{1}{2}\right) \cdot x.$$

In both cases the constants $\bar{c}_i$ and $c_i$ $(i \in \{0,1,2,3\})$ depend only on the distribution parameters but not on x or n. In each case the term with the $x \cdot e^x$ dependence prevails for large x. Therefore neither the Poisson nor the negative binomial distribution is fat tailed – as is well known.

However we find strong numerical evidence that the mixed Poisson – inverse – gamma distribution shows linear dependence between x and y for large x. In figure 1 the log-log plot for an expectation value of 0.001 and a variance of 1.01 times the expectation is compared for the Poisson – gamma and the Poisson – inverse – gamma case. The Poisson - gamma graph decreases very fast while the Poisson – inverse – gamma graph shows the typical linear dependence.

It is well known that the fat tail parameter $1/\xi$ is equal to the slope of the linear function y(x):

$$-\frac{1}{\xi} = \frac{dy}{dx}$$

We perform numerical calculations to evaluate $dy/dx$ over a huge range in parameter space of the mixed Poisson – inverse – gamma distribution and can summarise our findings as follows:

1) The function y(x) is linear for large x
2) $\frac{dy}{dx} \approx -s-1 \Rightarrow \frac{1}{\xi} \approx s+1$

Our results covering a wide range in parameter space are documented in table 1. Of course this is no rigorous mathematical proof that the first derivative is constant, but it gives a strong numerical evidence for this thesis. We additionally checked that $1/\xi$ stays constant within the interval 10 < x < 13 and not only at the end points.

Besides its relevance for this article the mixed Poisson – inverse gamma distribution is useful in insurance practice for modelling the claim numbers in a portfolio with storm exposure. Such portfolios show fat tailed claim number distributions. The extremely high total claim amounts are caused by the claim numbers and not by the severity of single claims. This parameter Poisson parameter (mixed by the inverse – gamma pdf) can be interpreted as the severity of the storm in two ways: Firstly as the physical power of the storm and secondly as the local concentration of insured goods (risk exposure) affected by the storm.

## 2.3 The Concept of Resolution

In the framework of classic BMS the number of claims delivered by the contract leads to the classification into a risk class and as a consequence to the premium amount. We are interested in the question if the forecast of future claim numbers via the past experience is accurate enough for classification and maybe a premium increase. If not, punishment of customers is not justified on a sound risk measuring basis and the customers may change to competing companies with a better, risk adjusted pricing system.

We therefore investigate if a well posed classification and distinction of risks is possible by the number of their observed claims. We define the "resolution" as the relation between the change in the conditional expectation value of claim number when $n_1$ changes to $n_1+1$ and the standard deviation of the conditional pdf. Formally this can be written as:

$$\mathrm{Re}\,soluion := \frac{E(N_2 = n_2 | N_1 = n_1 + 1) - E(N_2 = n_2 | N_1 = n_1)}{\sqrt{Var(N_2 = n_2 | N_1 = n_1)}} \qquad (2.3.1)$$

If the resolution takes values of 1 or higher, we are in a "high resolution" regime. In this case the expectations of two neighbouring conditional pdfs are separated at least by a distance comparable to their widths. Therefor the two pdfs can be recognised as two distinct ones. In other words: If the resolution of the two conditional pdfs (for $n_1$ and $n_1+1$) is high, the overlap between them is small.
In the Poisson – gamma case the resolution evaluates to

$$\mathrm{Re}\,soluion = \sqrt{\frac{J_2}{(n_1 + \alpha)(\beta + J_1 + J_2)}} \qquad (2.3.2)$$

as follows from (2.1.7). This means that the resolution can reach a value near 1 only if: $n_1 = 0 \land \alpha, \beta < 1$. We can also learn from (2.3.2) that in the favourable case $\alpha, \beta < 1$ there are only two "satisfactorily" resolved groups: Namely $n_1 = 0$ and $n_1 > 0$. For all other cases the resolution is rather low. Many of the examples presented in the literature lie within this parameter regime of low resolution, see for instance (Mack 1997, chap. 2.5.2. with α = 1,1001; β = 6,458) or (Bühlmann 2005, chap. 2.4 with α = 1,107; β = 7,67).
For the Poisson – inverse – gamma case an analytic representation of the resolution is still possible. It follows from equation (2.2.9). Discussion of this is however not as easy as that of (2.3.2).

## 3. Results and Outlook

In comparison to the classical Poisson – gamma model we introduced a new Bonus – Malus model with the inverse gamma mixing distribution. We include the calculation of the conditional distributions into our consideration. The benefits of the new model and the new pricing concept are the following:

- The Poisson – inverse gamma model fits quite well to real world claim frequency data in lines of business that have not been treated in literature up to now.
- Risk adequate premium principles can be included in our concept, because we look at the pdf and not only on the first moment.
- We find non linear (no credibility formula!) premium increase with the number of experienced claims in the case of the Poisson – inverse gamma model. Therefore small numbers of experienced claims are less penalized than large ones.
- We find a relatively smaller premium increase in cases where the conditional distribution has a wider tail compared to that of the standard model. The risk of "accidental" claim occurrence but not belonging to a worse risk group is taken into consideration here.

We further introduce the concept of resolution. This quantity helps to decide in which cases (dependent on parameter values!) the number of experienced claims is a reasonable risk – classification variable.

Our results show that there is a danger of anti - selection of insurance portfolios in a competitive market, because unjustified punishment is very likely if expectation value principle is used for pricing. This problem exists in a wide region of the parameter space. Especially an insurance company that introduces a BMS as a forerunner in the market is in danger to loose contracts because of a premium increase (after claim occurrence) that is not consistent with the risk adequate price.

We identify two sources of problems leading to anti - selection:

The mixing distribution of the model does not coincide with the "real world" portfolio's mixing distribution. One possible reason for this inefficiency can be distortions of the real world mixing distribution induced by large scale risk volumes within the portfolio. The other source is the use of an inadequate premium principle.

In an efficient financial market risk adequate pricing will be superior to a pure expectation value principle in the long run.

**Acknowledgement**
The Author gratefully acknowledges discussions on the subject of this paper with K.D. Schmidt and M. Zocher, TU Dresden.

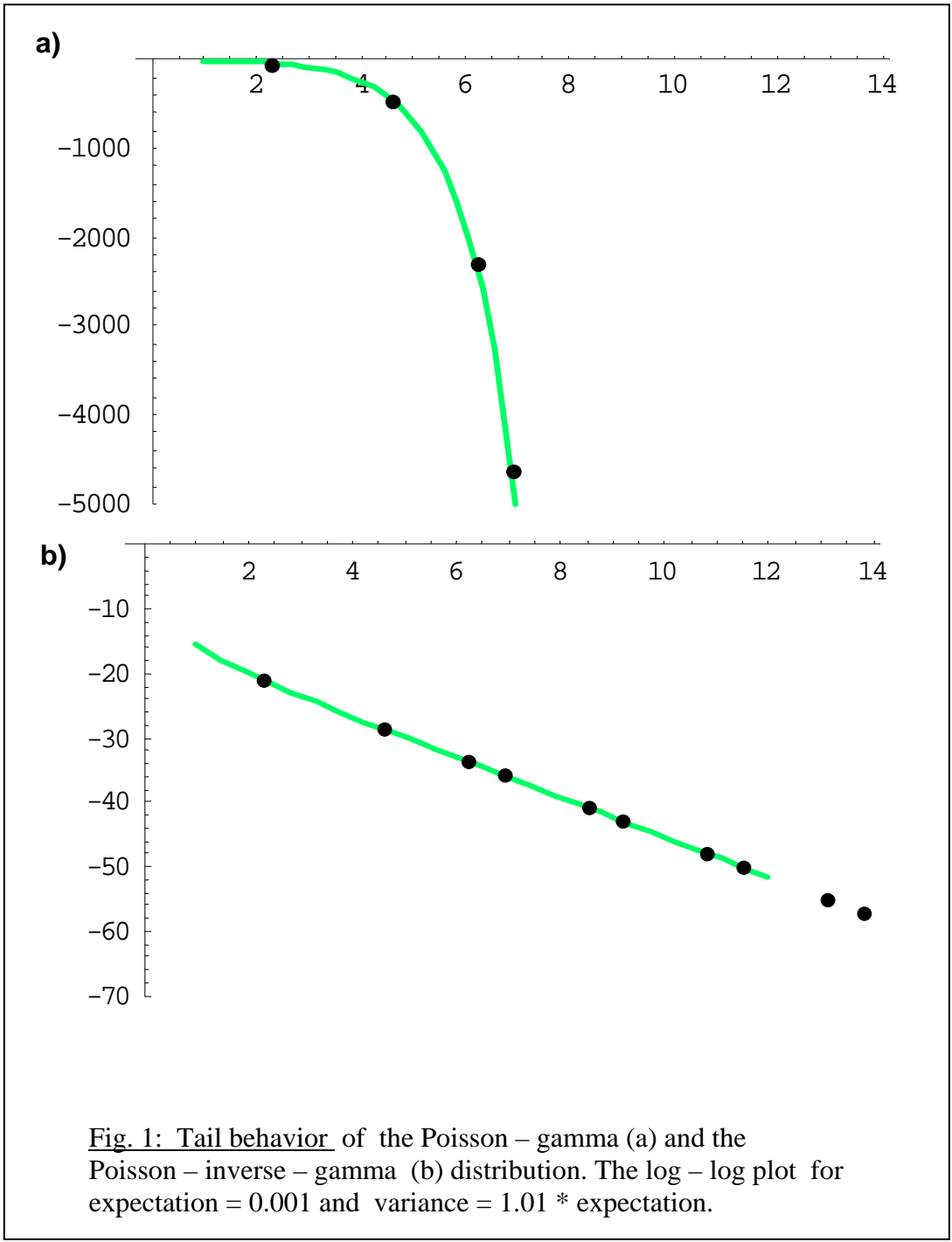

Fig. 1: Tail behavior of the Poisson – gamma (a) and the Poisson – inverse – gamma (b) distribution. The log – log plot for expectation = 0.001 and variance = 1.01 * expectation.

| m | s | ew | var | $1/\xi$ (x =10) | $1/\xi$ (x =13) |
|---:|---:|---:|---:|---:|---:|
| 0,0011 | 2,100E+00 | 0,001 | 0,00101 | -3,100E+00 | -3,109E+00 |
| 0,00101 | 2,010E+00 | 0,001 | 0,0011 | -3,010E+00 | -3,020E+00 |
| 0,001001 | 2,001E+00 | 0,001 | 0,002 | -3,001E+00 | -3,011E+00 |
| 0,001000111 | 2,000E+00 | 0,001 | 0,01 | -3,000E+00 | -3,011E+00 |
| 0,00100001 | 2,000E+00 | 0,001 | 0,1 | -3,000E+00 | -3,010E+00 |
| 0,001000001 | 2,000E+00 | 0,001 | 1 | -3,000E+00 | -3,010E+00 |
| 0,0001 | 1,100E+00 | 0,001 | infinity | -2,100E+00 | -2,110E+00 |
| 0,001 | 2,000E+00 | 0,001 | infinity | -3,000E+00 | -3,010E+00 |
| 0,02 | 3,000E+00 | 0,01 | 0,0101 | -4,000E+00 | -4,009E+00 |
| 0,011 | 2,100E+00 | 0,01 | 0,011 | -3,100E+00 | -3,109E+00 |
| 0,0101 | 2,010E+00 | 0,01 | 0,02 | -3,010E+00 | -3,019E+00 |
| 0,010011111 | 2,001E+00 | 0,01 | 0,1 | -3,001E+00 | -3,010E+00 |
| 0,01000101 | 2,000E+00 | 0,01 | 1 | -3,000E+00 | -3,010E+00 |
| 0,0100001 | 2,000E+00 | 0,01 | 10 | -3,000E+00 | -3,009E+00 |
| 0,001 | 1,100E+00 | 0,01 | infinity | -2,100E+00 | -2,110E+00 |
| 0,01 | 2,000E+00 | 0,01 | infinity | -3,000E+00 | -3,009E+00 |
| 1,1 | 1,200E+01 | 0,1 | 0,101 | -1,300E+01 | -1,301E+01 |
| 0,2 | 3,000E+00 | 0,1 | 0,11 | -4,000E+00 | -4,008E+00 |
| 0,11 | 2,100E+00 | 0,1 | 0,2 | -3,100E+00 | -3,108E+00 |
| 0,101111111 | 2,011E+00 | 0,1 | 1 | -3,011E+00 | -3,020E+00 |
| 0,10010101 | 2,001E+00 | 0,1 | 10 | -3,001E+00 | -3,010E+00 |
| 0,10001001 | 2,000E+00 | 0,1 | 100 | -3,000E+00 | -3,009E+00 |
| 0,01 | 1,100E+00 | 0,1 | infinity | -2,100E+00 | -2,109E+00 |
| 0,1 | 2,000E+00 | 0,1 | infinity | -3,000E+00 | -3,009E+00 |
| 101 | 1,020E+02 | 1 | 1,01 | -1,032E+02 | -1,030E+02 |
| 11 | 1,200E+01 | 1 | 1,1 | -1,300E+01 | -1,301E+01 |
| 2 | 3,000E+00 | 1 | 2 | -4,000E+00 | -4,008E+00 |
| 1,111111111 | 2,111E+00 | 1 | 10 | -3,111E+00 | -3,119E+00 |
| 1,01010101 | 2,010E+00 | 1 | 100 | -3,010E+00 | -3,018E+00 |
| 1,001001001 | 2,001E+00 | 1 | 1000 | -3,001E+00 | -3,009E+00 |
| 0,1 | 1,100E+00 | 1 | infinity | -2,100E+00 | -2,108E+00 |
| 1 | 2,000E+00 | 1 | infinity | -3,000E+00 | -3,008E+00 |
| 10010 | 1,002E+03 | 10 | 10,1 | -1,026E+03 | -1,004E+03 |
| 1010 | 1,020E+02 | 10 | 11 | -1,032E+02 | -1,030E+02 |
| 110 | 1,200E+01 | 10 | 20 | -1,300E+01 | -1,301E+01 |
| 21,11111111 | 3,111E+00 | 10 | 100 | -4,111E+00 | -4,119E+00 |
| 11,01010101 | 2,101E+00 | 10 | 1000 | -3,101E+00 | -3,108E+00 |
| 10,1001001 | 2,010E+00 | 10 | 10000 | -3,010E+00 | -3,017E+00 |
| 1 | 1,100E+00 | 10 | infinity | -2,100E+00 | -2,108E+00 |
| 10 | 2,000E+00 | 10 | infinity | -3,000E+00 | -3,007E+00 |
| 1211,111111 | 1,311E+01 | 100 | 1000 | -1,406E+01 | -1,411E+01 |
| 516,6666667 | 6,167E+00 | 100 | 2500 | -7,144E+00 | -7,172E+00 |
| 201,010101 | 3,010E+00 | 100 | 10000 | -4,001E+00 | -4,016E+00 |
| 10 | 1,100E+00 | 100 | infinity | -2,100E+00 | -2,107E+00 |
| 100 | 2,000E+00 | 100 | infinity | -2,996E+00 | -3,006E+00 |
| 11101,0101 | 1,210E+01 | 1000 | 100000 | -1,260E+01 | -1,308E+01 |
| 100 | 1,100E+00 | 1000 | infinity | -2,096E+00 | -2,106E+00 |
| 1000 | 2,000E+00 | 1000 | infinity | -2,955E+00 | -3,004E+00 |

Table 1: Linearity of the function y(x) and the value of the slope
-- scan in parameter space:
The parameters of the evaluated Poisson - inverse - gamma distributions m and s as well as their expectation (ew) and variance (var); the numerical values of dy/dx at x = 10 (n = 2.2 10^4) and x = 13 (n = 4.4 10^5). It can be seen that the relation $1/\xi = s+1$ holds quite well.